\documentclass[aps,reprint,superscriptaddress,
nofootinbib,showpacs,amsmath,amssymb]{revtex4-1}

\usepackage{latexsym,graphicx,slashed,bm,dcolumn}
\usepackage{hyperref}

\newcommand{\be}[1]{\begin{equation}\label{#1}}
\newcommand{\beq}{\begin{equation}}
\newcommand{\eeq}{\end{equation}}
\def\ee{\end{equation}}
\newcommand{\beqn}[1]{\begin{eqnarray}\label{#1}}
\newcommand{\eeqn}{\end{eqnarray}}

\newcommand{\mat}[4]{\left(\begin{array}{cc}{#1}&{#2}\\{#3}&{#4}
\end{array}\right)}

\newcommand\PM[1]{\begin{pmatrix}#1\end{pmatrix}}
\renewcommand{\to}{\rightarrow}

\def\ov{\overline}

\def\lsim{\raise0.3ex\hbox{$\;<$\kern-0.75em\raise-1.1ex
\hbox{$\sim\;$}}}
\def\gsim{\raise0.3ex\hbox{$\;>$\kern-0.75em\raise-1.1ex
\hbox{$\sim\;$}}}
\def\cal{\mathcal}

\def\cM{{\cal M}}

\def\eps{\varepsilon}

\newcommand{\Dm}{\Delta m}
\newcommand{\dm}{\delta m}

\renewcommand{\vec}[1]{ \boldsymbol{#1}}
\newcommand{\vect}[1]{\mbox{\boldmath$#1$}}

\def\rB{{\rm B}}
\def\rL{{\rm L}}

\begin{document}

\title{Neutron lifetime puzzle and  neutron -- mirror neutron oscillation   }

\author{Zurab~Berezhiani}
\email{E-mail: zurab.berezhiani@lngs.infn.it}
\affiliation{Dipartimento di Fisica e Chimica, Universit\`a di L'Aquila, 67100 Coppito, L'Aquila, Italy} 
\affiliation{INFN, Laboratori Nazionali del Gran Sasso, 67010 Assergi,  L'Aquila, Italy}

%\date{}

\begin{abstract} 
The discrepancy between the  neutron lifetimes 
%$\tau_{\rm trap}$ and $\tau_{\rm beam}$  
measured  in the beam and trap experiments can be explained via the neutron $n$  conversion 
into mirror neutron $n'$,  its dark partner from  parallel mirror sector,  
provided that $n$ and $n'$ have a tiny mass splitting order $10^{-7}$~eV. 
%$\vert m_n - m_{n'} \vert \sim 10^{-7}$~eV.   
In large magnetic fields used in beam experiments $n-n'$ transition is resonantly enhanced and 
can transform of about a per cent fraction of neutrons into mirror neutrons which decay in   
invisible mode. 
%which decay in invisible channel produces mirror particles, $n' \to p' e' \bar\nu'_e$, 
Thus less protons will be produced and 
the measured value $\tau_{\rm beam}$ appears larger than $\beta$-decay time $\tau_{\beta} = \tau_{\rm trap}$. 
Some phenomenological and astrophysical consequences of this scenario are also briefly discussed.  
  
\end{abstract}

\maketitle

%%%%%%%%%%%%%%%%%%%%%%%%%%%%%%%%%%%
%%%     END Abstract 
%%%%%%%%%%%%%%%%%%%%%%%%%%%%%%%%%%%%%%

%\bigskip 
%\bigskip 

\medskip 

\noindent {\bf 1.} 
Exact determination of the neutron lifetime remains a problem. It is measured in two types of experiments. 
The trap experiments measure the disappearance rate of the ultra-cold neutrons (UCN)  
 counting the survived UCN after storing them for different times  in material or magnetic traps, 
 and  determine the neutron decay width $\Gamma_n = \tau_n^{-1}$.  
The beam experiments are the appearance experiments, 
measuring the width of $\beta$-decay $n\to pe\bar\nu_e$, $\Gamma_\beta=\tau_\beta^{-1}$,  
by counting the produced protons  in the monitored beam of cold neutrons. 
As far as in the Standard Model (SM) the neutron decay always   produces  a proton,  
both methods should measure the same value, $\Gamma_n = \Gamma_\beta$.  
However, as it was pointed out in Refs. \cite{Serebrov:2011},  
the tension is mounting between the results obtained by two methods. 
At present, the experimental results using the trap 
\cite{Mampe:1993,Serebrov:2005,Pichlmaier:2010,Steyerl:2012, Arzumanov:2015,Serebrov:2017,Ezhov:2014,Pattie:2017} and the beam  \cite{Byrne:1996,Yue:2013} methods separately yield  
\begin{eqnarray} 
&& \tau_{\rm trap} = 879.4 \pm 0.5~{\rm s}  \label{trap}\\ 
&& \tau_{\rm beam} = 888.0 \pm 2.0~{\rm s} \label{beam}
\eeqn
with the discrepancy  of about $4\sigma$: 
$ \Delta\tau = \tau_{\rm beam}  - \tau_{\rm trap} = (8.6 \pm 2.1)$~s. 
Barring the possibility of uncontrolled systematic errors and considering the problem as real, 
then a new physics  must be invoked which could consistently explain the relations between  
the decay width $\Gamma_n$, $\beta$-decay rate  $\Gamma_\beta$,   
and the measured values (\ref{trap}) and (\ref{beam}). 

Some time ago I proposed   a way out  \cite{INT}
assuming that the neutron has a new decay channel $n \to n' X$ into a `dark neutron' $n'$  
and some light bosons $X$ among which a photon, due to a mass gap $m_{n} - m_{n'}  \simeq 1$~MeV  
(see also \cite{Fornal}).  
Then the beam and trap methods would measure correspondingly the neutron $\beta$-decay rate   
 $\Gamma_\beta = \tau_{\rm beam}^{-1}$  and the total width
$\Gamma_n = \Gamma_\beta + \Gamma_{\rm new} = \tau_{\rm trap}^{-1} $,  
so that $\tau_{\rm trap}/\tau_{\rm beam}$ discrepancy between (\ref{trap}) and (\ref{beam}) could be explained  
by a branching ratio $\Gamma_{\rm new}/\Gamma_n \simeq 0.01$. 

However, as it was argued recently in Ref. \cite{f2}, such a solution is disfavored by 
recent experiments  \cite{Mund,UCNA} that measured $\beta$-asymmetry parameter
using different techniques    (the cold and ultra-cold neutrons respectively).  
Their results  are in perfect agreement and 
determine the axial current coupling  $g_A$ with 
%impressive 
one per mille  precision: 
\be{gA}
g_A = 1.2755 \pm 0.0011\, .
\ee 
%On the other hand, 
In the SM frames  $\tau_\beta$ and $g_A$  are related as 
\be{CMS}
\tau_\beta (1+3g_A^2) = (5172.0 \pm 1.1)~{\rm s}
\ee  
which relation is essentially free from the uncertainties related to radiative corrections \cite{f2}. 
Then, for $g_A$ in the range (\ref{gA}),  Eq. (\ref{CMS}) predicts the neutron $\beta$-decay time  
\be{tauSM}
\tau_\beta^{\rm SM} = 879.5 \pm 1.3~{\rm s}  
\ee
perfectly agreeing with the value of $\tau_{\rm trap}$ (\ref{trap}) 
whereas in the dark decay scenario one expects  $ \tau_{\rm trap} \!< \!\tau_\beta\!=\!\tau_{\rm beam}$.  
Other way around, for $\tau_\beta\!=\!\tau_{\rm beam}$ Eq. (\ref{CMS})   
would imply  $g_A = 1.2681 \pm 0.0017$, more than $3.5\sigma$ away from (\ref{gA}). 
Hence, the dark decay  solution in fact 
replaces $\Delta\tau$ discrepancy by $g_A$ inconsistency  \cite{f2}. 
The situation does not improve neither by allowing  additional non-standard operators    
involving scalar or tensor currents in $\beta$-decay, 
and  $\tau_{\rm beam}/\tau_\beta$ incompatibility remains persistent \cite{BBB}.  

In the present letter I propose a $g_A$-consistent solution in which   
$\tau_{\rm trap}\!=\!\tau_\beta \!< \!\tau_{\rm beam} $.   
I assume that there exists a parallel/mirror hidden sector as a duplicate of our particle sector, 
so that all known particles: the electron $e$, proton $p$, neutron $n$, etc., 
have the mass-degenerate dark twins: $e'$, $p'$, $n'$, etc. (for review see Refs. \cite{Alice}). 
No fundamental principle forbids to our neutral particles, elementary as neutrinos 
or composite as the neutron, to have mixings with their mirror partners. 
Then $\tau_\beta^{\rm SM}/\tau_{\rm beam}$ discrepancy 
can be explained via neutron--mirror neutron mixing \cite{BB-nn'} 
which phenomenon is similar, and perhaps complementary \cite{BM}, 
to a baryon number violating  ($\Delta \rB\! =\!2$) mixing between the neutron and antineutron
\cite{Phillips}.  
But, in contrast to the latter, $\Delta \rB\!=\!1$ transition $n\!\to \!n'$  
 is not severely restricted by existing experimental bounds and can be rather effective.

\medskip 

\noindent 
{\bf 2.} 
Consider a theory $G_{\rm SM} \times G'_{\rm SM}$ with two gauge sectors   
where $G_{\rm SM}$ stands for the SM of ordinary (O) particles  
and $G'_{\rm SM}$ for its duplicate describing mirror (M) particles. 
The identical forms of their Lagrangians 
%${\cal L}$ and ${\cal L}'$ 
can be ensured by discrete $Z_2$ symmetry ${\rm SM}\leftrightarrow {\rm SM}'$
under which all O particles (fermions, Higgs and gauge bosons) exchange places with their 
M partners (`primed' fermions, Higgs and gauge bosons). 
If  $Z_2$ is exact,  then all M particles  should be exactly degenerate in mass with their O twins. 

%One can introduce also the mixed Lagrangian terms ${\cal L}_{\rm mix}$ 
%describing the possible 

There can exist also some feeble interactions between O and M  particles, e.g. in the form of  
effective $\rL$-violating operators $\frac{1}{M} l \phi l'\phi'$ 
which induce ``active-sterile" mixing between
our neutrinos $\nu_{e,\mu,\tau}$ and mirror neutrinos $\nu'_{e,\mu,\tau}$ \cite{ABS}.  
As for the mixing between the neutron and ``sterile" M neutron, $\eps \, \ov{n} n' + $ h.c.,  
it can be induced by TeV scale operators   $\frac{1}{\cM^5}(\ov{u} \ov{d} \ov{d}) (u'd'd') $ 
with quarks $u,d$ and mirror quarks $u',d'$  \cite{BB-nn'}. 
It violates $\rB$ and $\rB'$ separately but conserves the combination $\rB+\rB'$. 
Then, modulo $O(1)$ coefficients  depending on the operator structures, one has  
\be{eps} 
\eps \sim \frac{\Lambda_{\rm QCD}^6}{\cM^5} \sim\left(\frac{1~{\rm TeV}}{\cM}\right)^5 \times 10^{-10}~{\rm eV} \, .
\ee

One can envisage a situation when $Z_2$ is spontaneously broken e.g. 
 a scalar field $\eta$ which is odd under $Z_2$ symmetry, $\eta \to -\eta$, 
and couples to  O and M Higgses as $\lambda\eta (\phi^\dagger\phi - \phi^{\prime\dagger}\phi')$  \cite{BDM}.  
Then its non-zero VEV 
%coupling to O and M Higgses  $\lambda\eta (\phi^\dagger\phi - \phi^{\prime\dagger}\phi')$  
gives different contributions to mass terms  of $\phi$ and $\phi'$  in the Higgs potential 
and thus induces the difference between the VEVs of the latter. 
If the coupling $\lambda$ is small, then $Z_2$ breaking can be tiny, 
say $\langle \phi'\rangle/\langle\phi\rangle = 1 + O(10^{-13})$ or so. 
As far as the Yukawa couplings in two sectors are equal,  
then O and M quarks and leptons will get slightly different masses. 

So, let us consider that $n$ and $n'$ have a tiny mass splitting  $\Delta m = m_{n} - m_{n'} \sim 10^{-7}$~eV
which can be positive or negative 
(Cf. the neutron mass itself is measured with the precision of few eV.)
%The similar difference can exist between the O and M proton and electron, with unnoticeable difference 
%between their atomic physics.  
With mass gap being so small, $n-n'$ transition is not effective for destabilizing the nuclei \cite{BB-nn'}, 
but it will affect $n-n'$ oscillation pattern for free neutrons. 
In particular, the limits of Refs. \cite{Experiments} 
from experimental search of $n-n'$ oscillation obtained by assuming $\Dm=0$ are no more strictly applicable. 

\medskip 

\noindent 
{\bf 3.} 
Evolution of $n-n'$ system is described  Schr\"odinger equation $i d\Psi/dt = H\Psi$ 
where $\Psi = (\psi^+_n, \psi^-_n, \psi^+_{n'}, \psi^-_{n'})$ stands for wavefunctions of $n$ and $n'$ 
components  in two ($\pm$) polarization states. 
In background free vacuum conditions $4\times4$ Hamiltonian has the form $H=H_0 + H_{\rm dec}$: 
\be{H0} 
H_0 = \mat{\frac{\Dm}{2}   }{\eps}{\eps}{-\frac{\Dm}{2}  } , \quad H_{\rm dec} = 
- \frac{i}{2} \mat{\Gamma_\beta }{0}{0}{ \Gamma'_\beta} \, .  
\ee 
The average mass  of $n$ and ${n'}$ is omitted 
%$\frac12(m_n+m_{n'})$ 
since for $n-n'$ oscillation only the mass difference $\Delta m$ is relevant. 
%which can be positive or negative. 
%but  hereafter for the definiteness we take $\Dm > 0$. In addition, 
One can also set $\Gamma'_\beta = \Gamma_\beta$ neglecting 
a tiny difference between the decay rates of $n$ and $n'$. 

As far as we are interested in average oscillation probabilities,   it is convenient to consider the evolution 
in the basis of mass eigenstates where $H_0$ becomes diagonal:    
\be{mass-eigen} 
\psi_1^\pm = c_0 \psi_n^\pm + s_0 \psi_{n'}^\pm\, , \quad \psi_{2}^\pm = -s_0 \psi_n^\pm + c_0 \psi_{n'}^\pm \, ,
\ee 
with $c_0 = \cos\theta_0$ and  $s_0 = \sin\theta_0$, $\theta_0$ being $nn'$ mixing angle in vacuum 
which is the same for  both $\pm$ polarization states, 
$\tan 2\theta_0 = 2\eps/\Delta m$. 
In this way one takes into account also possible decoherence effects 
in $n-n'$ oscillation since the mass eigenstates do not oscillate but just propagate independently.  
The physical sense is transparent: producing  a neutron $n$ with $\pm$ polarization  is equivalent 
to producing mass eigenstates $\psi^\pm_1$ and $\psi_2^\pm$ 
respectively with probabilities $c_0^2$ and $s_0^2$.  
 Since $\psi^\pm_1$ interact as $n$ or $n'$ respectively with probabilities $c_0^2$ and $s_0^2$, 
and $\psi^\pm_2$ interact as $n$ or $n'$  with probabilities $s_0^2$ and $c_0^2$,
then the average probability  of finding $n$ after a time $t$ 
is $P_{nn} = c_0^4 + s_0^4 = 1 - \frac12 \sin^2 2\theta_0$,   
and that of finding $n'$ is 
\be{Pnn'} 
P_{nn'} = 1- P_{nn} = \frac12 \sin^2 2\theta_0 = 2 \frac{\eps^2}{\dm^2} \,   .
\ee 
Here  $\dm =\Dm \sqrt{1+ (2\eps/\Dm)^2}=\Dm/\cos2\theta_0$ is 
the mass gap between the eigenstates (\ref{mass-eigen}).  
As far as $\eps \ll \Dm$, we have $\dm \approx \Dm$, $\cos\theta_0 \approx1$ and 
$\sin\theta_0\approx \theta_0\approx \eps/\Dm$. 
In addition, since in real experimental situations the neutron free flight time between interactions is 
small, $t \ll \tau_\beta$, we have neglected the neutron decay  and corresponding 
overall factor $\exp(-\Gamma_\beta t)$ in these probabilities. 

The presence of matter background  and magnetic fields introduces an additional term in the 
Hamiltonian: 
\be{HI} 
% i \frac{d\psi}{dt} = H \psi ,  \quad \quad 
 H_I = \mat{V_n + \mu_n \vect{B}\vect{\sigma} }{0}{0}{V'_n + \mu'_n \vect{B}'\vect{\sigma} }  
\ee
which includes  the optical potentials $V_n,V'_n$ induced by O and M matter,  
and  interactions with respective magnetic fields $\vect{B}$ and $\vect{B}'$  \cite{BB-nn'}. 
Here $\vec{\sigma}=(\sigma_x,\sigma_y,\sigma_z)$ are the Pauli matrices, 
$\mu_n=-1.913 \mu_N = 6.031 \times 10^{-8}$~eV/T is  the neutron magnetic moment, 
and $\mu_{n'}\approx \mu_n$  is that of mirror neutron.  
In the following we neglect the presence, if any,  of M matter and M magnetic field at the Earth. 
In addition, since the neutron experiments are performed in perfect vacuum 
conditions, we neglect also 
%the O matter potential  
$V_n$. 

In uniform magnetic field $\vect{B}$ the spin quantization axis can be taken as 
the direction of $\vect{B}$,   and the Hamiltonian $H=H_0+H_I$ acquires a simple form
\be{mat44}
H = \PM{  \frac{\Dm}{2} - \Omega_B & 0 & \eps & 0 \\
0 &    \frac{\Dm}{2}  + \Omega_B  & ~ 0 ~ & ~ \eps ~ \\
\eps & 0 &  ~ - \frac{\Dm}{2} ~  & ~ 0 ~\\
0 & \eps  & ~ 0 ~ &  ~ - \frac{\Dm}{2} ~ }   
\ee
where $\Omega_B  = | \mu_n B | = (B/1\, {\rm T}) \times 60.31$~neV.  
In this case the Hamiltonian eigenstates are: 
\be{H-eigen} 
\psi_{1B}^\pm = c_B^\pm \psi_n^\pm + s_B^\pm \psi_{n'}^\pm\, , \quad 
\psi_{2B}^\pm = -s_B^\pm \psi_n^\pm + c_B^\pm \psi_{n'}^\pm
\ee 
with $c_B^\pm = \cos\theta_B^\pm$ and  $s_B^\pm= \sin\theta_B^\pm$. 
But now $nn'$ mixing angles  $\theta_{B}^\pm$ depend on polarization: 
\be{tan2B}
\tan 2\theta_{B}^{\pm} = \frac{2\eps}{\Delta m \mp \Omega_B} \, .
\ee
%\
Hence, in large magnetic fields, when $\Omega_B$ becomes comparable with  $\Dm$, 
one of the oscillation probabilities $P_{nn'}^{\pm} = \frac12 \sin^2 2\theta_B^\pm$ 
($+$ or $-$ depending on the sign of $\Dm$)  will be resonantly amplified,  
a phenomenon resembling the famous MSW effect in the neutrino oscillations. 

\medskip 

\noindent
{\bf 4.}  Trap experiments store an initial number of the UCN,   
count the amount of  neutrons survived for different times  $t$
and determine their disappearance rate $\Gamma_{\rm st}$ 
via exponential fit  $N_{\rm surv}(t)/N_{\rm in} = \exp(-\Gamma_{\rm st} t)$. 
%$N_{\rm surv}(t_1)/N_{\rm surv}(t_2) = \exp \big[- (t_1-t_2)/\tau_{\rm st} \big]$.  
In real experimental conditions there are always  some additional losses, 
and one has  to accurately estimate and substract their rates
for finding the true decay time,   $\tau_n^{-1} = \Gamma_{\rm st} - \Gamma_{\rm loss}$. 

These losses are dominated by the UCN absorption or up-scattering 
at the wall collisions, with a rate given by a product 
of the mean loss probability  per wall scattering $P$ 
and the mean frequency of scatterings $f$ averaged over the UCN velocity spectrum in the trap, 
$\Gamma_{\rm wall} = \langle P f \rangle$.  
It is controlled by measuring  $\Gamma_{\rm st}$  
for different frequencies $f$,  using traps of different sizes and varying the UCN velocities. 
In this way, one can determine $\tau_n=\tau_{\rm trap}$ by extrapolating the measured  values 
$\Gamma_{\rm st}$ to zero-scattering limit, also finding the neutron loss factor $P$. 

In the experiments with material traps the magnetic field is negligibly small, 
$\Omega_B \ll \Dm$, and $n-n'$ conversion probability is given by Eq. (\ref{Pnn'}). 
Per each wall collision the neutron would escape the trap with a probability 
$P_{nn'} =\frac12 \sin^22\theta_0 \simeq 2\theta_0^2$ which however 
should be included in the measured loss factor $P$. 
In particular, in experiment \cite{Serebrov:2005}  it was estimated as 
$P\simeq 2\times 10^{-6}$ (see also Ref. \cite{Pokotilovski} for more details).
%(Cf.  from quantum mechanical calculations 
%based on neutron optics  one would expect $P \simeq 3\times 10^{-7}$ \cite{Pokotilovski}).   
This gives a conservative  upper limit on $nn'$ mixing angle,  
%$P_{nn'} =2\theta_0^2 < 2\times 10^{-6$, 
$\theta_0 < 10^{-3}$ or so. 
%$\eps < 10^{-3} \dm$.  

Let us remark that this limit strictly applies if the mass difference $m_{n'} - m_n=-\Dm$ 
%$-\Dm$ between $\psi_1$ and $\psi_2$ states 
is less than the (positive) potential $V_n$ confinining neutrons in the trap. 
The Latter depends on the wall coating material, and for Fomblin Oil 
used in experiment \cite{Serebrov:2005} it is about $100$~neV.   
For $m_{n'} - m_n>100$~neV  the trapped UCN can be only in the lighter eigenstates $\psi_1^\pm$, 
and so the larger values of $\theta_0$ are can also also allowed.  This could contribute 
to anomalous UCN losses in the materials with higher potentials (e.g. $V_n=240$ neV for Beryllium) 
origin of which remains unclear in the context of neutron optics calculations  \cite{Serebrov:2004}.  
E.g. taking $\Dm \simeq -200$~neV and  $\theta_0 \simeq  3 \times 10^{-3}$, 
we get $P_{nn'} \simeq 2 \times 10^{-5}$,  close to the measured  loss factor 
for beryllium traps.  

The situation is somewhat different for magnetic traps. E.g. experiment \cite{Pattie:2017} 
uses a trap constructed as a Halbach array of permanent magnets with a surface field of 
about 1~T and additional externally applied holding field $B_\perp \sim 10^{-2}$~T,  
confining only $-$ polarized neutrons.  The true lifetime $\tau_n$ is assumed 
to be in practice equal to  the measured $\Gamma_{\rm st}^{-1}$,  
corrected by small $\Gamma_{\rm loss}$  dominated  by microphonic heating (0.23~s). 
The UCN losses on walls is inferred to occur 
 via the neutron depolarization which is effectively controlled by varying the holding 
field $B_\perp$ and gives less than 0.01~s correction. 
However,  the possibility of the losses due to $n-n'$ conversion is not taken into account, 
due to which per each wall scattering the UCN could escape with a probability 
$P_{nn'}^- = \frac12 \sin^2 2\theta_B^-$. 
For $\Dm >0$ the non-zero magnetic field can only suppress this probability,  
$P_{nn'}^- \simeq 2\theta_0^2/(1+ \Omega_B/\Dm)^2 < 2\theta_0^2$. 
For negative $\Dm$ above $-60$~neV or so, 
this probability can be resonantly enhanced in the vicinity 
of walls causing too big losses, but e.g. for $\Dm \simeq - 200$ neV this effect will be negligible. 
The role of $n-n'$ conversion in magnetic traps deserves a careful analysis, but generically 
one can expect the measured value $\tau_{\rm st}$ to be  less than true $\tau_n$.

Interestingly, experiments with the material  
\cite{Mampe:1993,Serebrov:2005,Pichlmaier:2010,Steyerl:2012, Arzumanov:2015,Serebrov:2017} 
and magnetic \cite{Ezhov:2014,Pattie:2017} traps 
yield somewhat different results, $\tau_{\rm mat} = 880.2 \pm 0.5$~s and $\tau_{\rm magn} = 877.8 \pm 0.7$~s. 
It is perhaps premature to consider this  discrepancy of about $2.7 \sigma$, 
$\tau_{\rm mat} - \tau_{\rm magn} = 2.4 \pm 0.9$~s,  as real 
but in principle it can naturally occur in our scenario if  $\theta_0 \simeq 10^{-3}$ or so. 

\medskip 
\noindent
{\bf 5.}  
As discussed in the introduction, the neutron `total' lifetime $\tau_n$ measured in the trap experiments (\ref{trap}) 
perfectly agrees the Standard Model prediction for $\beta$-decay (\ref{tauSM}), 
$\tau_{\rm trap} = \tau_\beta^{\rm SM}$. This in fact gives an upper limit on the rate of 
neutron dark decay \cite{INT,Fornal}, 
and in any case disfavors it as explanation of the neutron lifetime puzzle. 
Hence, the question remains: once $\tau_n$ 
is indeed the same as $\tau_\beta$, why then the measurements of the latter  
 in beam experiments \cite{Byrne:1996,Yue:2013} gives contradictory result  
with $\tau_{\rm beam}$ (\ref{beam}) of about one percent larger than $\tau_\beta$? 
There are two possibilities: either some fraction of protons produced in the trap  
is lost by yet unknown reasons, or in large magnetic fields  
($B=5$~T and 4.6~T  respectively in beam experiments \cite{Byrne:1996,Yue:2013}) 
some fraction of neutrons transforms into M neutrons then decaying 
via dark channel as $n'\to p'e'\bar\nu_e'$, 
and exactly this is the fraction missing the detection. 

Let us discuss the beam experiments (described in details in Refs. {\cite{Yue:2013}) 
also taking into account the effect of $n-n'$ oscillation. 
Their principal  scheme is shown in Fig. \ref{fig:beam}.  
The narrow beam of cold neutrons passes through the proton trap. 
%(with a diameter 8.4 mm) with large axial magnetic field  
At any moment the number of neutrons in the trap is   
$N_n = P^{\rm tr}_{nn}L \int_A da \int dv I(v)/v $ and the number of M neutrons is 
$N_{n'} = P^{\rm tr}_{nn'}L \int_A da \int dv I(v)/v $,  
where $A$ is the beam cross-sectional area, $L$ is the effective length of the trap,  
$I(v)$ is the velocity dependent fluence rate and $P^{\rm tr}_{nn} = 1 - P^{\rm tr}_{nn'}$ 
is the average survival probability of the neutron in the trap. 
Then the  count rate of protons produced by $\beta$-decay $n\to pe\bar\nu_e$  inside the trap  is
\be{Np} 
%\dot{N}^{\rm count}_p = frac{dN^{\rm count}_p}{dt}  
%\dot{N}_p= \frac{e_p L }{\tau_n}   \int_A da \int dv \frac{I(v)}{v}
\dot{N}_p= e_p  \Gamma_{\beta}  P^{\rm tr}_{nn} L   \int_A da \int dv \frac{I(v)}{v} \, , 
\ee 
$e_p$ being the counting efficiency. 
After passing the proton trap, beam hits the neutron counter, 
which is $^6$LiF  foil, and  the reaction products of neutron absorption by $^6$Li, 
alphas and tritons, are detected with  a net count rate 
\be{Nn} 
\dot{N}_{\alpha}  = 
%\frac{dN_{\alpha+t}^{\rm count}}{dt}  = 
e_\alpha \bar{v}  P^{\rm det}_{nn} \int_A da \int dv \frac{I(v)}{v} \, ,
\ee 
where $ e_\alpha$ is the counting efficiency 
normalized to the neutrons with a velocity $\bar{v} = 2200$ m/s, 
and $P^{\rm det}_{nn} = 1 - P^{\rm det}_{nn'}$ is the 
neutron survival probability at the position of the neutron detector. 
Hence, by taking the ratio of (\ref{Np}) and (\ref{Nn}), 
in reality one measures not $\tau_\beta$ but the value 
\be{tau-beam-new} 
\tau_{\rm beam}   =
 \left(\frac{e_p L }{e_\alpha \bar{v}}\right) \left(\frac{ \dot{N}_{\alpha} } {\dot{N}_{p} } \right) 
= \frac{P_{nn}^{\rm det}}{P_{nn}^{\rm tr}} \, \tau_{\beta} \, .
\ee
Thus, a per cent discrepancy between the measured value $\tau_{\rm beam}$ (\ref{beam}) 
and the SM predicted  $\tau_\beta$  (\ref{tauSM}) 
can be understood provided that $P_{nn}^{\rm tr}/P_{nn}^{\rm det} \simeq 0.99$, 
or $P_{nn'}^{\rm tr} - P_{nn'}^{\rm det} \simeq 10^{-2}$.  

%%%%%%%%%%%%%%%%%%%%%%%%%%%%%%%%%%%%%%%%%%%%
%%%%%%%%%%%%%%%%%%%%%%%%%%%%%%%%%%%%%%%%%%%%%
\begin{figure}[t]
\includegraphics[width=0.45\textwidth]{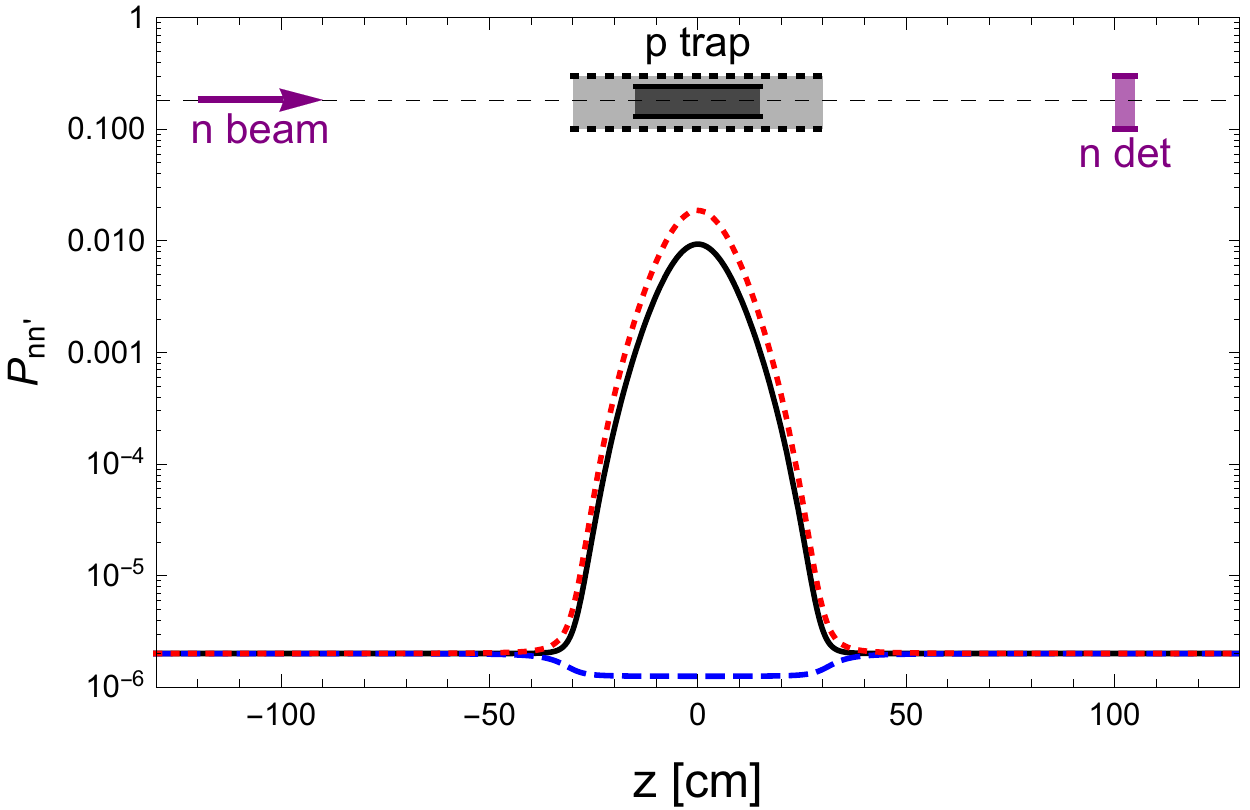} \\
\vspace{2mm}
\includegraphics[width=0.45\textwidth]{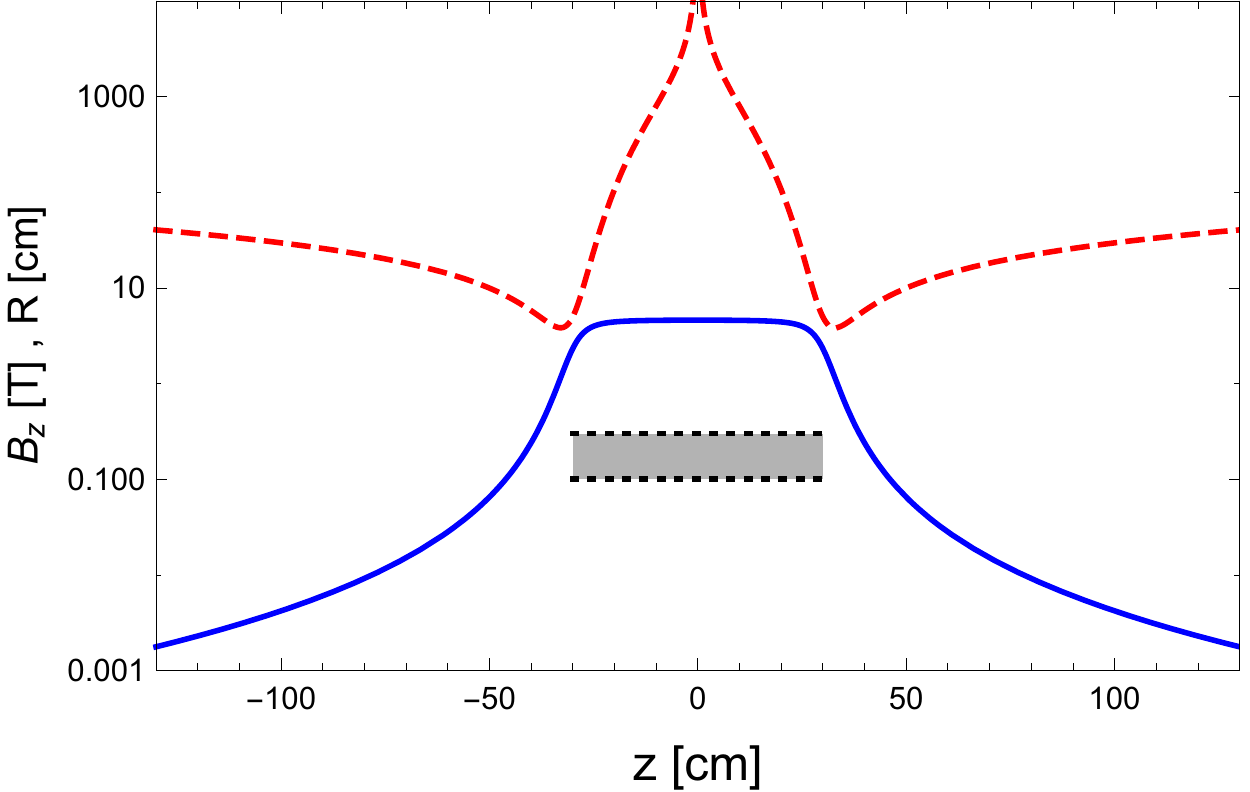}
\caption{{\it Upper panel:} The cold neutron beam passes through the proton trap 
located inside a solenoid (grey box)  inducing magnetic field with the central value $B_{\rm tr} =4.6$~T, 
and then hits the neutron detector.  
The red dotted, blue dashed and black solid curves respectively show the evolution of $P_{nn'}^\pm(z)$ 
and  their average $P_{nn'}(z)$  
%$P_{nn'} = \frac12\big(P_{nn'}^+ + P_{nn'}^- \big)$. 
for the parameters chosen as $\theta_0 = 10^{-3}$ and $\Dm = 280$~neV.  
{\it Lower panel:} Profile of the axial magnetic field $B(z)$ induced by continuous solenoid 
(with length 60 cm and diameter 10 cm) at the position $z$ from its centre. 
The dashed red curve shows the resonance length scale 
$R = B\big(dB(z)/dz\big)^{-1}$ as a function of $z$.
}
\label{fig:beam}
\end{figure}
%%%%%%%%%%%%%%%%%%%%%%%%%%%%%%%%%%%%%%%%%%%
%%%%%%%%%%%%%%%%%%%%%%%%%%%%%%%%%%%%%%%%%%%%

For determining the conversion probabilities $P_{nn'}^{\rm tr}$ and $P_{nn'}^{\rm det}$, 
one has to consider the propagation in a variable magnetic field. 
The field profile induced by a prototype continuous solenoid is shown in Fig. \ref{fig:beam}. 
Inside the trap  it is $B_{\rm tr}=4.6$~T, quickly falling outside the solenoid. 
Neutrons are born in small magnetic field and oscillate initially with 
$P_{nn'}^{\rm in} \simeq 2\theta_0^2$.  
Then they enter the trap where the field is large and $nn'$ mixing angles for $\pm$ polarizations 
become  (\ref{tan2B}). 
If evolution of  the wavefunction is adiabatic, the mass eigenstates 
$\psi^{\pm}_{1}$ and $\psi^{\pm}_{2}$ (\ref{mass-eigen}) would evolve correspondingly 
into the ``magnetic" eigenstates  $\psi^{\pm}_{1B}$   and $\psi^{\pm}_{2B}$ (\ref{H-eigen})   
which are detectable as $n$ respectively with the probabilities 
$c_{B_\pm}^2$ and $s_{B_\pm}^2$. 
Thus, the respective survival probabilities at the coordinate $z$ are fixed by 
the magnetic field value $B(z)$,  
$P_{nn}^{\pm}(z) = c_0^2  (c_{B}^{\pm})^2 +  s_0^2  (s_{B}^{\pm})^2$. 
Correspondingly,  $n-n'$ conversion probabilities are  
\be{Pnn'-adiab} 
P_{nn'}^{\pm}(z)  = 1- P_{nn}^{\pm}(z) = \frac12  - \frac12 \cos2\theta_0 \cos 2\theta_{B}^{\pm}(z)  
\ee 
where 
\be{cos2B}
\cos 2\theta_{B}^\pm = \frac{\cos 2\theta_0 (1\mp \frac{\Omega_B}{\Dm}\big)}
{\sqrt{\cos^2 2\theta_0 (1\mp \frac{\Omega_B}{\Dm}\big)^2 + \sin^2 2\theta_0} } \, .
 \ee
The evolution of $P_{nn'}^\pm(z)$  is shown in Fig. \ref{fig:beam} 
for $\theta_0 = 10^{-3}$ and $\Dm = 280$~neV. 
In this case the evolution is indeed adiabatic, as it can be directly checked by numerically solution 
of the evolution equation  which gives exactly the same result as Eq. (\ref{Pnn'-adiab}). 
The resonance is not crossed, but in the trap 
the value $\Omega_B = \vert \mu_n B \vert$ approaches $\Dm$ with about a per cent precision, 
and $n-n'$ conversion probability is strongly amplified for $+$ polarization state, 
$P_{nn'}^+  (z=0) \approx  \frac12 (1-\cos 2\theta_B^+) \approx 0.02$. 
Since the neutrons are unpolarized, one should average between two polarizations, 
$P_{nn'} = \frac12(P_{nn'}^+ +P_{nn'}^-)$,  getting $P_{nn'}^{\rm tr} \approx 0.01$. 
At the neutron detector magnetic field is again small and   
so $P_{nn'}^{\rm det} \approx P_{nn'}^{\rm in} \approx 2\theta_0^2 \approx 2\times 10^{-6}$.  
Then Eq. (\ref{tau-beam-new}) gives $\tau_{\rm beam}/\tau_\beta \approx 1 + P_{nn'}^{\rm tr} \approx 1.01$. 
Needless to say, for $\Dm<0$ the resonant amplification would occur instead for $-$ polarized neutrons 
but the average probability would remain the same. Thus the sign of $\Dm$ is irrelevant.  

The situation is even more interesting when $\Dm < 260$~neV and  the 
neutron crosses the resonance  before entering the proton trap, 
at some position $z_{\rm res}$ at which $B_{\rm res}  = B(z_{\rm res})= \vert \Dm/\mu_n \vert 
= (\Dm/100~{\rm nev}) \times 1.66$~T. 
Eq. (\ref{cos2B}) tells that $\cos2\theta^+_B$ vanishes when $B=B_{\rm res}$  
and it becomes negative at $B>B_{\rm res}$.
The $-$ polarization states $\psi_1^-$ and $\psi_2^-$  still evolve adiabatically  
respectively into $\psi_{1B}^-$ and $\psi_{2B}^-$,  with $\cos2\theta^+_B \approx 1$ 
and thus $P^-_{nn'}(z) \leq 2\theta_0^2$ at any position. 
But the evolution of  $\psi_1^+$ and $\psi_2^+$ is no more adiabatic  
and one has to take into account the Landau-Zener probability 
that at the resonance crossing  the state $\psi_1^+$ can jump into $\psi_{2B}^+$. 
The goodness of adiabaticity depends on parameter 
$\xi = \Dm \sin^2 2\theta_0 v^{-1} R(z_{\rm res}) $,  
where $v$ is the neutron velocity. The function 
$R(z) = \big(d \ln B/dz)^{-1}$ (shown in lower panel of Fig. \ref{fig:beam}) describes  
the resonance length scale, and it is typically $\sim10$~cm for $B_{\rm res} \sim 1$~T. 
Then at coordinates $z$ inside the trap we have
\be{non-adiab} 
P^+_{nn'}(z) = \frac12 - \left(\frac12 - e^{-\pi\xi/2} \right) \cos2\theta_0 \cos 2\theta_{B}^+(z) 
\ee 
The adiabatic limit (\ref{Pnn'-adiab}) corresponds to $\xi\gg 1$. 
However, in our case $\xi \ll 1$, so that $\exp(-\pi\xi/2) \approx 1 -\frac12 \pi\xi$.  
In addition, Eq. (\ref{cos2B}) tells that for $B_{\rm tr} - B_{\rm res} > 10^{-2}$~T or so 
one can take $\cos2\theta^+_B \approx -1$. 
Thus, the conversion probability averaged between $\pm$ polarizations becomes   
\beqn{final-num}
&& P^{\rm tr}_{nn'}  \approx \frac12  P_{nn'}^{+}(z=0) \approx \frac{\pi}{4}\xi   \nonumber \\ 
&& \simeq 10^{-2}  \left(\frac{2~{\rm km/s}}{v}\right) \left(\frac{\theta_0}{10^{-3}}\right)^2 
\left(\frac{B_{\rm res} }{1~{\rm T}}\right) 
 \left(\frac{ R_{\rm res} }{10~{\rm  cm}}\right) 
\eeqn
just in the range needed for explaining a one per cent difference between $\tau_{\rm beam}$ and $\tau_\beta$. 
Let us recall also that e.g. for $B_{\rm res}=1 \div 4$~T, corresponding to  $\Dm =60\div 240$~neV,  
the resonance length scale $R_{\rm res} = R(z_{\rm res})$ falls in the range of few cm 
almost independently of the inferred solenoid sizes. (Unfortunately, the detailed descriptions 
of the magnetic fields used in beam experiments \cite{Byrne:1996,Yue:2013} are not available, 
but the profile shown in Fig. \ref{fig:beam} is rather similar to that of Fig. 13
in Ref. \cite{Yue:2013}.)  

In future experiments $n-n'$ conversion can be rendered more adiabatic.  
One can increase the resonance length scale $R_{\rm res}$ by 1-2 orders of magnitude 
by constructing  magnetic fields with smooth enough profile.  
%slowly changing from the lower to larger values,  and thus achieve $P_{nn'}^+\approx 1$. 
Then spectacular effect can be expected: in the proton trap almost all neutrons of one polarization 
will be lost and almost all neutrons of other polarization will survive. So only a half of 
the initial neutrons will produce protons and the measured $\tau_{\rm beam}$ can appear 
twice as big as $\tau_\beta$.

\medskip

\noindent
{\bf 6.} 
Our scenario suggests interesting connection  between the neutron lifetime  and dark matter puzzles. 
Mirror atoms, invisible in terms of ordinary photons but gravitationally coupled  to our matter, 
can constitute a reasonable fraction of cosmological dark matter or even its entire amount. 
M baryons represent  a sort of asymmetric dark matter,
and its dissipative character can have  specific implications  for the cosmological evolution, 
formation and structure of galaxies and stars, etc. \cite{BCV} and for dark matter direct detection \cite{Cerulli}.  
Interestingly, the same $\rB\!-\!\rL$ (and CP) violating interactions between O and M particles that 
that induce $\nu-\nu'$ or $n-n'$ mixings,  can induce baryon asymmetries 
in both O and M worlds in the early universe and naturally explain the 
dark and visible matter fractions,  $\Omega_{\rB'}/\Omega_{\rB} \simeq 5$ \cite{BB-PRL}. 
There can be some common interactions between two sectors, e.g. with the gauge bosons 
of the flavor symmetry which can induce oscillation effects between O and M 
neutral Kaons, etc. which picture also suggests interesting realizations of minimal flavor violation \cite{MFV}. 
As for $n-n'$ mixing itself, it can have intriguing effects on  
ultra-high energy cosmic rays propagating at cosmological distances \cite{nn'-cosm}. 
Its implications for the neutron stars which can be slowly transformed 
in mixed O-M neutron stars, with a maximal mass and radii by a factor 
of $\sqrt 2$ lower than that of ordinary ones, were briefly discussed in \cite{INT} 
and will be analysed in details elsewhere \cite{Massimo}. It also is tempting to consider 
the possibility that $n-n'$ conversion has some effect in neutron rich heavy unstable nuclides 
and can be somehow related to the reactor neutrino anomaly \cite{Serebrov:2018}. 

Some additional remarks are in order. We assumed that mass splitting between 
ordinary and mirror neutrons, $\Dm = m_{n'} - m_n \sim 10^{-7}$~eV, is induced by a tiny breaking 
of mirror $Z_2$ symmetry. Then the same order mass differences 
can be expected also between O and M protons and electrons, etc. 
but microphysics of two sectors should be essentially the same. 
There is nothing wrong in this possibility, 
and it might be also related to the necessity of asymmetric post-inflationary 
reheating between O and M sectors \cite{BDM}.  However, there is also a tempting possibility that 
$Z_2$ is exact and $\Dm=0$, but instead the order $10^{-7}$~eV difference  
between potentials $V_n$ and $V'_n$  in (\ref{HI})   effectively emerges
due to environmental reasons.  
One can consider some long range 5th forces, 
with radii comparable to the Earth radius or solar system size,  related to 
e.g. light baryophoton interactions in each sector \cite{ABK}, or to 
the difference of graviton/dilaton coupling between O and M components 
e.g. in the context of bigravity theories \cite{bigravity}.  In first case the force induced by the Earth 
is repulsive for the neutron which is equivalent of having $\Dm>0$, while in the second case it 
would be attractive and equivalent to $\Dm<0$. This splitting can be effective 
at the Earth whereas somewhere in cosmological voids it could be vanishingly small.

We considered the effects of $n-n'$ mass mixing $\eps$ given in (\ref{eps}),
induced by effective $\Delta\rB=1$ interactions between O and M quarks in the context of some 
new physics, as e.g. seesaw mechanism in Ref. \cite{BB-nn'}.   
Generically this underlying physics  should violate also CP-invariance, 
and in principle it can induce interactions with the electromagnetic field \cite{Arkady},  
$\mu_{nn'} F_{\mu\nu} \ov{n} \sigma^{\mu\nu} n'$ and $d_{nn'} F_{\mu\nu} \ov{n} \sigma^{\mu\nu}\gamma^5 n'$ 
(and equivalent terms with $F_{\nu\nu} \to F'_{\nu\nu}$), where $\mu_{nn'}$ and $d_{nn'}$ 
respectively are the transitional magnetic  moment and 
electric dipole moment between $n$ and $n'$.  
%which can be of the same order. 
Both of these transitional moments can have interesting effects 
to be studied in details \cite{Variano}, especially the CP-violating one $d_{nn'}$, 
also because in beam experiments the large electric fields are also used.

To summarize, we discussed a scenario based on $n-n'$ conversion which can be effective 
in large magnetic fields, and can resolve the neutron lifetime puzzle 
explaining why the beam and trap experiments get different results. 
In addition, it suggests that the lifetimes measured in material and magnetic traps 
can be somewhat different, and it can also shed some more light 
on the origin of the UCN anomalous losses in material traps. 
Effects for the neutron propagation in matter depend on the sign of $\Dm$ 
and deserve careful study. 
%potentially explain the anomalous neutron losses in material traps with high surface potentials. 
If our proposal  is correct, this would mean that installations  
used in the beam experiments are in fact effective machines  that transform the neutrons 
in dark matter. 
This can be easily tested experimentally by varying the magnetic field profiles and rendering 
$n-n'$ conversion more adiabatic. 
%in magnetic fields with a smooth profile slowly changing from the lower to larger values. 
%In this case very significant fraction of the neutrons in the beam, up to 50 \%,  can disappear
%which can result in $\tau_{\rm beam} \simeq 1500$~s.   
%Other tests can be done by making use of large magnetic fields
In particular, such tests can be done  in planned 30~m baseline experiment 
searching for $n\to n'$ transition and $n \to n' \to n$ regeneration \cite{nn'-exp} 
which is under construction at the HFIR reactor of the Oak Ridge National Laboratory.

\bigskip 

{\bf Acknowledgements} 
\medskip

\noindent
I thank R. Biondi, Y. Kamyshkov, Y. Pokotilovsky and A. Serebrov for  help 
and useful information.


\begin{thebibliography}{99}

%%%%%%%%%%%%%%%%%%%%%%  Zurab 

%\cite{Serebrov:2011re}
\bibitem{Serebrov:2011} 
 A.~Serebrov and A.~Fomin, Phys. Procedia {\bf 17}, 19 (2011);  
% A.~P.~Serebrov and A.~K.~Fomin, Physics Procedia {\bf 17}, 199 (2011); 
  %``New evaluation of neutron lifetime from UCN storage experiments and beam experiments,''
% [arXiv:1104.4238 [nucl-ex]]; 
  %%CITATION = ARXIV:1104.4238;%%
 % \bibitem{Greene} 
G.~L.~ Greene and P. Geltenbort, Sci. Am. {\bf 314}, 36 (2016).  
%doi:10.1038/scientificcamerican0416-36. 
 
 %\cite{Mampe:1993an}
\bibitem{Mampe:1993} 
  W.~Mampe {\it et al.}, 
  % L.~N.~Bondarenko, V.~I.~Morozov, Y.~N.~Panin and A.~I.~Fomin,
  %``Measuring neutron lifetime by storing ultracold neutrons and detecting inelastically scattered neutrons,''
  JETP Lett.\  {\bf 57}, 82 (1993).
%  [Pisma Zh.\ Eksp.\ Teor.\ Fiz.\  {\bf 57}, 77 (1993)].
  %%CITATION = JTPLA,57,82;%%
 
 %\cite{Serebrov:2004zf}
\bibitem{Serebrov:2005} 
  A.~P.~Serebrov {\it et al.},
  %``Measurement of the neutron lifetime using a gravitational trap and a low-temperature Fomblin coating,''
  Phys.\ Lett.\ B {\bf 605}, 72 (2005);  \,
 % doi:10.1016/j.physletb.2004.11.013
 %  [nucl-ex/0408009]; 
  %see also 
  %%CITATION = doi:10.1016/j.physletb.2004.11.013;%%
%\cite{Serebrov:2007ve}
%\bibitem{Serebrov:2007ve} 
%  A.~P.~Serebrov {\it et al.},
  %``Neutron lifetime measurements using gravitationally trapped ultracold neutrons,''
  Phys.\ Rev.\ C {\bf 78}, 035505 (2008).
  % [nucl-ex/0702009].
% doi:10.1103/PhysRevC.78.035505
%  [nucl-ex/0702009 [NUCL-EX]].
  %%CITATION = doi:10.1103/PhysRevC.78.035505;%% 
 
 %\cite{Pichlmaier:2010zz}
\bibitem{Pichlmaier:2010} 
  A.~Pichlmaier {\it et al.},
  %, V.~Varlamov, K.~Schreckenbach and P.~Geltenbort,
  %``Neutron lifetime measurement with the UCN trap-in-trap MAMBO II,''
  Phys.\ Lett.\ B {\bf 693}, 221 (2010).
 % doi:10.1016/j.physletb.2010.08.032
  %%CITATION = doi:10.1016/j.physletb.2010.08.032;%% 
  
%\cite{Steyerl:2012zz}
\bibitem{Steyerl:2012} 
  A.~Steyerl {\it et al.}, 
  % J.~M.~Pendlebury, C.~Kaufman, S.~S.~Malik and A.~M.~Desai,
  %``Quasielastic scattering in the interaction of ultracold neutrons with a liquid wall and application in a reanalysis of the Mambo I neutron-lifetime experiment,''
  Phys.\ Rev.\ C {\bf 85}, 065503 (2012).  
%  doi:10.1103/PhysRevC.85.065503
  %%CITATION = doi:10.1103/PhysRevC.85.065503;%%%\cite{Mampe:1989xx}
  
  %\cite{Arzumanov:2015tea}
\bibitem{Arzumanov:2015} 
  S.~Arzumanov {\it et al.},
  % L.~Bondarenko, S.~Chernyavsky, P.~Geltenbort, V.~Morozov, V.~V.~Nesvizhevsky, Y.~Panin and A.~Strepetov,
  %``A measurement of the neutron lifetime using the method of storage of ultracold neutrons and detection of inelastically up-scattered neutrons,''
  Phys.\ Lett.\ B {\bf 745}, 79 (2015). 
  % doi:10.1016/j.physletb.2015.04.021
  %%CITATION = doi:10.1016/j.physletb.2015.04.021;%%
%reevaluated results of 
 %  S. Arzumanov {\it  et al.}, Nucl. Instrum. Methods, Sect. A 611 (2009) 186.
 
  %\cite{Serebrov:2017bzo}
\bibitem{Serebrov:2017} 
%  A.~P.~Serebrov {\it et al.},
%  JETP Lett. {\bf 106}, 623 (2017)
  %``Neutron lifetime measurements with the big gravitational trap for ultracold neutrons,''
%  [arXiv:1712.05663 [nucl-ex]].
  %%CITATION = ARXIV:1712.05663;%% 
  A.~P.~Serebrov {\it et al.},
  %``Neutron lifetime measurements with the big gravitational trap for ultracold neutrons,''
  Phys.\ Rev.\ C {\bf 97}, 
  %no. 5, 
  055503 (2018).
%  doi:10.1103/PhysRevC.97.055503
%  [arXiv:1712.05663 [nucl-ex]].
  %%CITATION = doi:10.1103/PhysRevC.97.055503;%%

  %\cite{Ezhov:2014tna}
\bibitem{Ezhov:2014} 
  V.~F.~Ezhov {\it et al.}, JETP {\bf 107}, 11 (2018). 
  %``Measurement of the neutron lifetime with ultra-cold neutrons stored in a magneto-gravitational trap,''
%  [arXiv:1412.7434 [nucl-ex]].
  %%CITATION = ARXIV:1412.7434;%%

%\cite{Pattie:2017vsj}
\bibitem{Pattie:2017} 
  R.~W.~Pattie, Jr. {\it et al.},
  %``Measurement of the neutron lifetime using a magneto-gravitational trap and in situ detection,''
  Science {\bf 360}, no. 6389, 627 (2018).
%  doi:10.1126/science.aan8895
%  [arXiv:1707.01817 [nucl-ex]].
  %%CITATION = doi:10.1126/science.aan8895;%%   
  
 %\cite{Byrne:1996zz}
\bibitem{Byrne:1996} 
  J.~Byrne  {\it et al.}, 
  %and P.~G.~Dawber,
  %``A Revised Value for the Neutron Lifetime Measured Using a Penning Trap,''
  Europhys.\ Lett.\  {\bf 33}, 187 (1996). 
  % doi:10.1209/epl/i1996-00319-x
  %%CITATION = doi:10.1209/epl/i1996-00319-x;%% 
  
 %\cite{Yue:2013qrc}
\bibitem{Yue:2013} 
  A.~T.~Yue {\it et al.}, 
 % M.~S.~Dewey, D.~M.~Gilliam, G.~L.~Greene, A.~B.~Laptev, J.~S.~Nico, W.~M.~Snow and F.~E.~Wietfeldt,
  %``Improved Determination of the Neutron Lifetime,''
  Phys.\ Rev.\ Lett.\  {\bf 111}, 
  %no. 22, 
  222501 (2013);  
%  doi:10.1103/PhysRevLett.111.222501
%  [arXiv:1309.2623 [nucl-ex]].  
  %%CITATION = doi:10.1103/PhysRevLett.111.222501;%% 
  %revaluated results of 
%\cite{Nico:2004ie}
%\bibitem{Nico:2005} 
  J.~S.~Nico {\it et al.},
  %``Measurement of the neutron lifetime by counting trapped protons in a cold neutron beam,''
  Phys.\ Rev.\ C {\bf 71}, 055502 (2005). 
%  doi:10.1103/PhysRevC.71.055502
%  [nucl-ex/0411041]; 
  %%CITATION = doi:10.1103/PhysRevC.71.055502;%%
 

  \bibitem{INT} 
Z. Berezhiani, "Unusual effects in $n-n'$ conversion", 
talk at the Workshop INT-17-69W, Seattle, 23-27 Oct. 2017, 
\verb+http://www.int.washington.edu/talks/WorkShops/+ \verb+int_17_69W/People/Berezhiani_Z/Berezhiani3.pdf+ 
% See also  arXiv:1807.xxxxx [hep-ph]. 

%\cite{Fornal:2018eol}
\bibitem{Fornal} 
  B.~Fornal and B.~Grinstein,
  %``Dark Matter Interpretation of the Neutron Decay Anomaly,''
  Phys.\ Rev.\ Lett.\  {\bf 120}, 
  %no. 19, 
  191801 (2018)
%  doi:10.1103/PhysRevLett.120.191801
  [arXiv:1801.01124 [hep-ph]].
%  arXiv:1801.01124 [hep-ph].  
  %%CITATION = ARXIV:1801.01124;%%
    
%\cite{Czarnecki:2018okw}
\bibitem{f2} 
  A.~Czarnecki, W.~J.~Marciano and A.~Sirlin,
  %``Neutron Lifetime and Axial Coupling Connection,''
  Phys.\ Rev.\ Lett.\  {\bf 120}, 
  %no. 20, 
  202002 (2018) 
%  doi:10.1103/PhysRevLett.120.202002
[arXiv:1802.01804 [hep-ph]].
  %%CITATION = doi:10.1103/PhysRevLett.120.202002;%%
 
 %\cite{Mund:2012fq}
\bibitem{Mund} 
  D.~Mund {\it et al.}, 
%  B.~Maerkisch, M.~Deissenroth, J.~Krempel, M.~Schumann, H.~Abele, A.~Petoukhov and T.~Soldner,
  %``Determination of the Weak Axial Vector Coupling from a Measurement of the Beta-Asymmetry Parameter A in Neutron Beta Decay,''
  Phys.\ Rev.\ Lett.\  {\bf 110}, 172502 (2013).
%  doi:10.1103/PhysRevLett.110.172502
%  [arXiv:1204.0013 [hep-ex]].
  %%CITATION = doi:10.1103/PhysRevLett.110.172502;%%
 
 %\cite{Brown:2017mhw}
\bibitem{UCNA} 
  M.~A.-P.~Brown {\it et al.}, 
  %[UCNA Collaboration],
  %``New result for the neutron $\beta$-asymmetry parameter $A_0$ from UCNA,''
  Phys.\ Rev.\ C {\bf 97}, 
  %no. 3, 
  035505 (2018).
%  doi:10.1103/PhysRevC.97.035505
%  [arXiv:1712.00884 [nucl-ex]].
  %%CITATION = doi:10.1103/PhysRevC.97.035505;%%
 
 \bibitem{BBB}
 B. Belfatto, R. Beradze and Z. Berezhiani, in preparation 
 
 \bibitem{Alice}
%\cite{Berezhiani:2003xm}
%\bibitem{Berezhiani:2003xm} 
  Z.~Berezhiani,
  %``Mirror world and its cosmological consequences,''
  Int.\ J.\ Mod.\ Phys.\ A {\bf 19}, 3775 (2004);   
%  doi:10.1142/S0217751X04020075
%  [hep-ph/0312335]; 
  %%CITATION = doi:10.1142/S0217751X04020075;%%
  %%CITATION = HEP-PH/0312335;%%
 %\cite{Berezhiani:2008zza}
%\bibitem{Berezhiani:2008zza} 
 %
%\cite{Berezhiani:2005ek}
%\bibitem{Berezhiani:2005ek}
%  Z.~Berezhiani, 
``Through the looking-glass: Alice's adventures in mirror world,''
 In  I. Kogan Memorial Volume {\it From Fields to Strings, 
  Circumnavigating Theoretical Physics}, World Scientific  (2005),  Eds. M. Shifman et al.,  vol. 3, pp. 2147-2195 
%  doi:10.1142/9789812775344_0055
% \verb+ doi:10.1142/9789812775344_0055+ 
[hep-ph/0508233]; 
 % hep-ph/0508233.
  %%CITATION = HEP-PH/0508233;%%
%\cite{Foot:2014mia}
%\bibitem{Foot} 
%Z.~Berezhiani,
  %``Unified picture of ordinary and dark matter genesis,''
  Eur.\ Phys.\ J.\ ST {\bf 163}, 271 (2008); 
 % doi:10.1140/epjst/e2008-00824-6
  %%CITATION = doi:10.1140/epjst/e2008-00824-6;%% 
   %%CITATION = 00619,163,271;%%
 R.~Foot,
  %``Mirror dark matter: Cosmology, galaxy structure and direct detection,''
  Int.\ J.\ Mod.\ Phys.\ A {\bf 29}, 1430013 (2014).  
 % doi:10.1142/S0217751X14300130
%  [arXiv:1401.3965 [astro-ph.CO]].
  %%CITATION = doi:10.1142/S0217751X14300130;%%
 
 \bibitem{BB-nn'}
 %\cite{Berezhiani:2005hv}
  Z.~Berezhiani and L.~Bento,
  %``Neutron - mirror neutron oscillations: How fast might they be?,''
  Phys.\ Rev.\ Lett.\  {\bf 96}, 081801 (2006);  
 % doi:10.1103/PhysRevLett.96.081801
 % [hep-ph/0507031].  
  %%CITATION = doi:10.1103/PhysRevLett.96.081801;%%
   %%CITATION = HEP-PH/0507031;%%
     %\cite{Berezhiani:2006je}
%\bibitem{Berezhiani:2006je} 
 % Z.~Berezhiani and L.~Bento,
  %``Fast neutron: Mirror neutron oscillation and ultra high energy cosmic rays,''
  Phys.\ Lett.\ B {\bf 635}, 253 (2006); 
 % doi:10.1016/j.physletb.2006.03.008
%  [hep-ph/0602227]; 
  %%CITATION = doi:10.1016/j.physletb.2006.03.008;%%
%
 %\cite{Berezhiani:2008bc}
%\bibitem{More} 
  Z.~Berezhiani,
  %``More about neutron - mirror neutron oscillation,''
  Eur.\ Phys.\ J.\ C {\bf 64}, 421 (2009).
%  doi:10.1140/epjc/s10052-009-1165-1
%  [arXiv:0804.2088 [hep-ph]]; 
  %%CITATION = doi:10.1140/epjc/s10052-009-1165-1;%%

%\cite{Berezhiani:2015afa}
    \bibitem{BM} 
  Z.~Berezhiani,
  %``Neutron?antineutron oscillation and baryonic majoron: low scale spontaneous baryon violation,''
  Eur.\ Phys.\ J.\ C {\bf 76}, 
  %no. 12, 
  705 (2016). 
 % doi:10.1140/epjc/s10052-016-4564-0
%  [arXiv:1507.05478 [hep-ph]]; 
 %%CITATION = doi:10.1140/epjc/s10052-016-4564-0;%% 

  
 \bibitem{Phillips} 
 %\cite{Kuzmin:1970nx}
%\bibitem{Kuzmin:1970nx} 
  V.~Kuzmin,
  %``Cp violation and baryon asymmetry of the universe,''
 % Pisma Zh.\ Eksp.\ Teor.\ Fiz.\  {\bf 12}, 335 (1970).
JETP\ Lett.\  {\bf 12}, 335 (1970); 
  %%CITATION = ZFPRA,12,335;%%
%\cite{Mohapatra:1980qe}
%\bibitem{Mohapatra:1980qe} 
  R.~N.~Mohapatra and R.~E.~Marshak,
  %``Local B-L Symmetry of Electroweak Interactions, Majorana Neutrinos and Neutron Oscillations,''
  Phys.\ Rev.\ Lett.\  {\bf 44}, 1316 (1980); 
%  Erratum: [Phys.\ Rev.\ Lett.\  {\bf 44}, 1643 (1980)].
 % doi:10.1103/PhysRevLett.44.1644.2, 10.1103/PhysRevLett.44.1316
  %%CITATION = doi:10.1103/PhysRevLett.44.1644.2, 10.1103/PhysRevLett.44.1316;%%
  %\cite{Phillips:2014fgb}
 for reviews see D.~G.~Phillips {\it et al.},
  %``Neutron-Antineutron Oscillations: Theoretical Status and Experimental Prospects,''
  Phys.\ Rept.\  {\bf 612}, 1 (2016); 
%  doi:10.1016/j.physrep.2015.11.001
%  [arXiv:1410.1100 [hep-ex]]; 
  %%CITATION = doi:10.1016/j.physrep.2015.11.001;%%
 %\cite{Babu:2013yww}
%\bibitem{Babu:2013yww} 
  K.~S.~Babu {\it et al.},
  %``Neutron-Antineutron Oscillations: A Snowmass 2013 White Paper,''
  arXiv:1310.8593 [hep-ex].
  %%CITATION = ARXIV:1310.8593;%%


 \bibitem{ABS}
%\cite{Akhmedov:1992hh}
%\bibitem{Akhmedov:1992hh} 
  E.~Akhmedov, Z.~Berezhiani and G.~Senjanovic,
  %  E.~K.~Akhmedov, Z.~Berezhiani and G.~Senjanovic,
  %``Planck scale physics and neutrino masses,''
  Phys.\ Rev.\ Lett.\  {\bf 69}, 3013 (1992); 
%  doi:10.1103/PhysRevLett.69.3013
% [hep-ph/9205230]; 
  %%CITATION = doi:10.1103/PhysRevLett.69.3013;%%
  %\cite{Foot:1991py}
%\bibitem{Foot:1991py} 
  R.~Foot, H.~Lew and R.~Volkas,
  %``Possible consequences of parity conservation,''
  Mod.\ Phys.\ Lett.\ A {\bf 7}, 2567 (1992); 
 % doi:10.1142/S0217732392004031
  %%CITATION = doi:10.1142/S0217732392004031;%%
%\bibitem{BM}
%\cite{Foot:1995pa}
%\bibitem{Foot:1995pa} 
  R.~Foot and R.~Volkas,
  %``Neutrino physics and the mirror world: How exact parity symmetry explains the solar neutrino deficit, the atmospheric neutrino anomaly and the LSND experiment,''
  Phys.\ Rev.\ D {\bf 52}, 6595 (1995);  
%  doi:10.1103/PhysRevD.52.6595
 % [hep-ph/9505359];
  %%CITATION = doi:10.1103/PhysRevD.52.6595;%%
%\cite{Berezhiani:1995yi}
%\bibitem{Berezhiani:1995yi} 
  Z.~Berezhiani and R.~N.~Mohapatra,
  %``Reconciling present neutrino puzzles: Sterile neutrinos as mirror neutrinos,''
  Phys.\ Rev.\ D {\bf 52}, 6607 (1995).
%  doi:10.1103/PhysRevD.52.6607
%  [hep-ph/9505385].
  %%CITATION = doi:10.1103/PhysRevD.52.6607;%%
    %%CITATION = HEP-PH/9505385;%%

 \bibitem{BDM}
%\cite{Berezhiani:1995am}
%\bibitem{Berezhiani:1995am} 
  Z.~Berezhiani, A.~D.~Dolgov and R.~N.~Mohapatra,
  %``Asymmetric inflationary reheating and the nature of mirror universe,''
  Phys.\ Lett.\ B {\bf 375}, 26 (1996);  
%  doi:10.1016/0370-2693(96)00219-5
%  [hep-ph/9511221]; 
  %%CITATION = doi:10.1016/0370-2693(96)00219-5;%%
  %%CITATION = HEP-PH/9511221;%%
%
%\cite{Berezhiani:1996sz}
%\bibitem{Berezhiani:1996sz} 
Z.~Berezhiani,
  %``Astrophysical implications of the mirror world with broken mirror parity,''
Acta Phys.\ Polon.\ B {\bf 27}, 1503 (1996); 
% [hep-ph/9602326]. 
  %%CITATION = HEP-PH/9602326;%%
     %\cite{Mohapatra:2017lqw}
%\bibitem{Mohapatra:2017} 
  R.~N.~Mohapatra and S.~Nussinov,
  %``Constraints on Mirror Models of Dark Matter from Observable Neutron-Mirror Neutron Oscillation,''
  Phys.\ Lett.\ B {\bf 776}, 22 (2018).
%  doi:10.1016/j.physletb.2017.11.022
%  [arXiv:1709.01637 [hep-ph]].

\bibitem{Experiments}
%\cite{Ban:2007tp}
%\bibitem{Ban:2007tp}
  G.~Ban {\it et al.},
%  A Direct experimental limit on neutron -- mirror neutron oscillations,
 Phys.\ Rev.\ Lett.\  {\bf 99}, 161603 (2007);  
%  [arXiv:0705.2336 [nucl-ex]]; 
  %%CITATION = PRLTA,99,161603;%%
%\cite{Serebrov:2007gw}
%\bibitem{Serebrov1}
  A.~Serebrov {\it et al.},
%  Experimental search for neutron -- mirror neutron oscillations using storage of ultracold neutrons,
Phys.\ Lett.\  B {\bf 663}, 181 (2008);    
% [arXiv:0706.3600 [nucl-ex]]; 
%   Phys.\ Lett.\  B {\bf 663} (2008) 181  [arXiv:0706.3600 [nucl-ex]].
 %%CITATION = PHLTA,B663,181;%%
%\cite{Altarev:2009tg}
%\bibitem{Altarev}
  I.~Altarev {\it et al.},
%  Neutron to Mirror  Neutron Oscillations in the Presence of Mirror Magnetic Fields,
 Phys.\ Rev.\  D {\bf 80}, 032003 (2009);     
%  [arXiv:0905.4208 [nucl-ex]];   
%\cite{Bodek:2009zz}
%\bibitem{Bodek}
%  K.~Bodek {\it et al.},
 % Additional results from the dedicated search for neutron mirror neutron oscillations,
% Nucl.\ Instrum.\ Meth. \  A {\bf 611}, 141 (2009); 
  %%CITATION = NUIMA,A611,141;%%
 %\cite{Serebrov:2009zz}
%\bibitem{Serebrov2}
  A.~Serebrov {\it et al.},
%  Search for neutron mirror neutron oscillations in a laboratory experiment with ultracold neutrons,
 Nucl.\ Instrum.\ Meth. \  A {\bf 611}, 137 (2009);
 %   Nucl.\ Instrum.\ Meth.\  A {\bf 611} (2009) 137.
% [arXiv:0809.4902 [nucl-ex]]; 
  %%CITATION = NUIMA,A611,137;%%
%    Phys.\ Rev.\  D {\bf 80} (2009) 032003  [arXiv:0905.4208 [nucl-ex]].
 %%CITATION = PHRVA,D80,032003;%%
  % \bibitem{Nesti}
 %\cite{Berezhiani:2012rq}
%\bibitem{Berezhiani:2012rq} 
  Z.~Berezhiani and F.~Nesti,
  %``Magnetic anomaly in UCN trapping: signal for neutron oscillations to parallel world?,''
  Eur.\ Phys.\ J.\ C {\bf 72}, 1974 (2012); 
%  doi:10.1140/epjc/s10052-012-1974-5
%  [arXiv:1203.1035 [hep-ph]]. 
  %%CITATION = doi:10.1140/epjc/s10052-012-1974-5;%%
%\cite{Berezhiani:2017jkn}
%\bibitem{ILL} 
  Z.~Berezhiani {\it et al.}, arXiv:1712.05761 [hep-ex] 
  %, R.~Biondi, P.~Geltenbort, I.~A.~Krasnoshchekova, V.~E.~Varlamov, A.~V.~Vassiljev and O.~M.~Zherebtsov,
  %``New experimental limits on neutron - mirror neutron oscillations in the presence of mirror magnetic field,''
 (Eur.\ Phys.\ J.\ C -- in press). 
  %%CITATION = ARXIV:1712.05761;%% 
 

\bibitem{Pokotilovski}
Y. Pokotilovski, I. Natkaniec and K. Holderna-Natkaniec, 
Physica B {\bf 403}, 1942 (2008). 

%\cite{Serebrov:2004zg}
\bibitem{Serebrov:2004} 
  A.~Serebrov {\it et al.},
  %``UCN anomalous losses and the UCN capture cross-section on material defects,''
  Phys.\ Lett.\ A {\bf 335}, 327 (2005). 
%  doi:10.1016/j.physleta.2004.12.032
%  [nucl-ex/0408010].
  %%CITATION = doi:10.1016/j.physleta.2004.12.032;%%

 \bibitem{BCV}
%\cite{Berezhiani:2000gw}
%\bibitem{Berezhiani:2000gw} 
  Z.~Berezhiani, D.~Comelli and F.~L.~Villante,
  %``The Early mirror universe: Inflation, baryogenesis, nucleosynthesis and dark matter,''
  Phys.\ Lett.\ B {\bf 503}, 362 (2001); 
%  doi:10.1016/S0370-2693(01)00217-9
% [hep-ph/0008105].
  %%CITATION = doi:10.1016/S0370-2693(01)00217-9;%%
    %%CITATION = HEP-PH/0008105;%%
%\bibitem{BCV2}
%\cite{Ignatiev:2003js}
%bibitem{Ignatiev}
  A.~Y.~Ignatiev and R.~R.~Volkas,
  %``Mirror dark matter and large scale structure,''
  Phys.\ Rev.\ D {\bf 68},  023518 (2003); 
% [hep-ph/0304260]; 
  %%CITATION = HEP-PH/0304260;%%
%
%\cite{Berezhiani:2003wj}
%\bibitem{Berezhiani:2003wj} 
  Z.~Berezhiani, P.~Ciarcelluti, D.~Comelli and F.~L.~Villante,
  %``Structure formation with mirror dark matter: CMB and LSS,''
  Int.\ J.\ Mod.\ Phys.\ D {\bf 14}, 107 (2005); 
%  doi:10.1142/S0218271805005165
%  [astro-ph/0312605]; 
  %%CITATION = doi:10.1142/S0218271805005165;%%
  %%CITATION = ASTRO-PH/0312605;%%
%\cite{Berezhiani:2005vv}
%\bibitem{Berezhiani:2005vv} 
  Z.~Berezhiani, S.~Cassisi, P.~Ciarcelluti and A.~Pietrinferni,
  %``Evolutionary and structural properties of mirror star MACHOs,''
  Astropart.\ Phys.\  {\bf 24}, 495 (2006). 
%  doi:10.1016/j.astropartphys.2005.10.002
%  [astro-ph/0507153].
  %%CITATION = doi:10.1016/j.astropartphys.2005.10.002;%%  
  
%\cite{Cerulli:2017jzz}
\bibitem{Cerulli} 
  R.~Cerulli {\it et al.}, 
  %, P.~Villar, F.~Cappella, R.~Bernabei, P.~Belli, A.~Incicchitti, A.~Addazi and Z.~Berezhiani,
  %``DAMA annual modulation and mirror Dark Matter,''
  Eur.\ Phys.\ J.\ C {\bf 77}, 
  %no. 2, 
  83 (2017); 
  %doi:10.1140/epjc/s10052-017-4658-3
%  [arXiv:1701.08590 [hep-ex]]. 
  %%CITATION = doi:10.1140/epjc/s10052-017-4658-3;%%
%\bibitem{Addazi:2015cua} 
  A.~Addazi {\it et al.}, 
  %Z.~Berezhiani, R.~Bernabei, P.~Belli, F.~Cappella, R.~Cerulli and A.~Incicchitti,
  %``DAMA annual modulation effect and asymmetric mirror matter,''
  Eur.\ Phys.\ J.\ C {\bf 75}, 
  %no. 8, 
  400 (2015).
%  doi:10.1140/epjc/s10052-015-3634-z
%  [arXiv:1507.04317 [hep-ex]].
  %%CITATION = doi:10.1140/epjc/s10052-015-3634-z;%%. 

  

\bibitem{BB-PRL}
%\cite{Bento:2001rc}
%\bibitem{Bento:2001rc} 
  L.~Bento and Z.~Berezhiani,
  %``Leptogenesis via collisions: The Lepton number leaking to the hidden sector,''
  Phys.\ Rev.\ Lett.\  {\bf 87}, 231304 (2001); 
%  doi:10.1103/PhysRevLett.87.231304
% [hep-ph/0107281]; 
  %%CITATION = doi:10.1103/PhysRevLett.87.231304;%%
 %
 %\cite{Bento:2002sj}
%\bibitem{Bento:2002sj} 
%  L.~Bento and Z.~Berezhiani,
  %``Baryon asymmetry, dark matter and the hidden sector,''
  Fortsch.\ Phys.\  {\bf 50}, 489 (2002) 
%  doi:10.1002/9783527610853.ch8
  %%CITATION = doi:10.1002/9783527610853.ch8;%% 
 % 
 %\cite{Bento:2001nb}
%\bibitem{Bento:2001nb} 
%  L.~Bento and Z.~Berezhiani,
  %``Baryogenesis: The Lepton leaking mechanism,''
  [hep-ph/0111116];
  %%CITATION = HEP-PH/0111116;%%
%\cite{Berezhiani:2013paa}
%\bibitem{NOW2012} 
Z.~Berezhiani,
  %``Sterile Neutrinos and Leptogenesis of Matter and Dark Matter,''
  Nucl.\ Phys.\ Proc.\ Suppl.\  {\bf 237-238}, 263 (2013); 
 % doi:10.1016/j.nuclphysbps.2013.04.104
  %%CITATION = doi:10.1016/j.nuclphysbps.2013.04.104;%%
%
%\cite{Berezhiani:2016ong}
%\bibitem{ADM} 
 % Z.~Berezhiani,
 % ``Anti-dark matter: a hidden face of mirror world,''
  arXiv:1602.08599 [astro-ph.CO].
  %%CITATION = ARXIV:1602.08599;%%

%\cite{Berezhiani:1996ii}
\bibitem{MFV} 
  Z.~Berezhiani,
  %``Unified picture of the particle and sparticle masses in SUSY GUT,''
  Phys.\ Lett.\ B {\bf 417}, 287 (1998); 
%  doi:10.1016/S0370-2693(97)01359-2
 % [hep-ph/9609342].
  %%CITATION = doi:10.1016/S0370-2693(97)01359-2;%%
%\cite{Berezhiani:2001mh}
%\bibitem{Berezhiani:2001mh} 
  Z.~Berezhiani and A.~Rossi,
  %``Flavor structure, flavor symmetry and supersymmetry,''
  Nucl. Phys. Proc. Sup.\  {\bf 101}, 410 (2001). 
 % doi:10.1016/S0920-5632(01)01527-4
%  [hep-ph/0107054].
  %%CITATION = doi:10.1016/S0920-5632(01)01527-4;%%


\bibitem{nn'-cosm} 
 %\cite{Berezhiani:2011da}
%\bibitem{Berezhiani:2011da} 
  Z.~Berezhiani and A.~Gazizov,
  %``Neutron Oscillations to Parallel World: Earlier End to the Cosmic Ray Spectrum?,''
  Eur.\ Phys.\ J.\ C {\bf 72}, 2111 (2012);
%  doi:10.1140/epjc/s10052-012-2111-1
 % [arXiv:1109.3725 [astro-ph.HE]]; 
  %%CITATION = doi:10.1140/epjc/s10052-012-2111-1;%%  
  Z. Berezhiani, R. Biondi and A. Gazizov, in press. 
 
\bibitem{Massimo}
Z. Berezhiani, R. Biondi, M. Mannarelli and F. Tonelli, in preparation
 
 %\cite{Serebrov:2018mva}
\bibitem{Serebrov:2018} 
  A.~P.~Serebrov {\it et al.}, 
%  , R.~M.~Samoilov, I.~A.~Mitropolsky and A.~M.~Gagarsky,
  %``Neutron lifetime, dark matter and search for sterile neutrino,''
  arXiv:1802.06277 [nucl-ex].
  %%CITATION = ARXIV:1802.06277;%%


\bibitem{ABK} 
%\cite{Babu:2016rwa}
%\bibitem{Babu:2016rwa} 
  K.~S.~Babu and R.~N.~Mohapatra,
  %``Limiting Equivalence Principle Violation and Long-Range Baryonic Force from Neutron-Antineutron Oscillation,''
  Phys.\ Rev.\ D {\bf 94}, 
  %no. 5, 
  054034 (2016); 
 % doi:10.1103/PhysRevD.94.054034
%  [arXiv:1606.08374 [hep-ph]].
  %%CITATION = doi:10.1103/PhysRevD.94.054034;%%
  %%CITATION = doi:10.1140/epjc/s10052-016-4564-0;%% 
%\cite{Addazi:2016rgo}
%\bibitem{Addazi:2016rgo} 
  A.~Addazi, Z.~Berezhiani and Y.~Kamyshkov,
  %``Gauged $B-L$ number and neutron?antineutron oscillation: long-range forces mediated by baryophotons,''
  Eur.\ Phys.\ J.\ C {\bf 77}, 
  %no. 5, 
  301 (2017).
 % doi:10.1140/epjc/s10052-017-4870-1
%  [arXiv:1607.00348 [hep-ph]]; 
  %%CITATION = doi:10.1140/epjc/s10052-017-4870-1;%% 

  \bibitem{bigravity} 
   %\cite{Berezhiani:2009kv}
%\bibitem{Berezhiani:2009kv} 
 Z.~Berezhiani, F.~Nesti, L.~Pilo and N.~Rossi,
  %``Gravity Modification with Yukawa-type Potential: Dark Matter and Mirror Gravity,''
  JHEP {\bf 0907}, 083 (2009); 
%  doi:10.1088/1126-6708/2009/07/083
%  [arXiv:0902.0144 [hep-th]].
  %%CITATION = doi:10.1088/1126-6708/2009/07/083;%%
%\cite{Berezhiani:2009kx}
%  Z.~Berezhiani, F. Nesti, L.~Pilo and N.~Rossi,
  %``Mirror Matter, Mirror Gravity and Galactic Rotational Curves,''
  Eur.\ Phys.\ J.\ C {\bf 70}, 305 (2010); 
 % doi:10.1140/epjc/s10052-010-1457-5
 % [arXiv:0902.0146 [astro-ph.CO]].
  %%CITATION = doi:10.1140/epjc/s10052-010-1457-5;%%
     %\cite{Berezhiani:2007zf}
%\bibitem{Berezhiani:2007zf} 
see also  Z.~Berezhiani, D.~Comelli, F.~Nesti and L.~Pilo,
  %``Spontaneous Lorentz Breaking and Massive Gravity,''
  Phys.\ Rev.\ Lett.\  {\bf 99}, 131101 (2007).
%  doi:10.1103/PhysRevLett.99.131101
%  [hep-th/0703264 [HEP-TH]].
  %%CITATION = doi:10.1103/PhysRevLett.99.131101;%%
 

  
  %\cite{Berezhiani:2015uya}
\bibitem{Arkady} 
  Z.~Berezhiani and A.~Vainshtein,
  %``Neutron-Antineutron Oscillation as a Signal of CP Violation,''
  arXiv:1506.05096 [hep-ph].
  %%CITATION = ARXIV:1506.05096;%%

\bibitem{Variano}
Z. Berezhiani, R. Biondi, Y. Kamyshkov and L. Varriano, in preparation
  
  \bibitem{nn'-exp}
  %\cite{Broussard:2017yev}
%\bibitem{Broussard:2017yev} 
  L.~J.~Broussard {\it et al.},
  %``New Search for Mirror Neutrons at HFIR,''
  arXiv:1710.00767 [hep-ex]; 
  %%CITATION = ARXIV:1710.00767;%%
   %\cite{Berezhiani:2017azg}
%\bibitem{HFIR} 
 see also  Z.~Berezhiani {\it et al.}, 
%M.~Frost, Y.~Kamyshkov, B.~Rybolt and L.~Varriano,
  %``Neutron Disappearance and Regeneration from Mirror State,''
  Phys.\ Rev.\ D {\bf 96}, 
  %no. 3, 
  035039 (2017). 
 % doi:10.1103/PhysRevD.96.035039
%  [arXiv:1703.06735 [hep-ex]]; 
  %%CITATION = doi:10.1103/PhysRevD.96.035039;%%

   \end{thebibliography}
 \end{document}